\documentclass[journal]{new-aiaa}
\usepackage[utf8]{inputenc}
\usepackage{textcomp}
\usepackage{graphicx}
\usepackage{subfig} 
\usepackage{amsmath}
\usepackage[version=4]{mhchem}
\usepackage{siunitx}
\usepackage{longtable,tabularx}
\usepackage{multirow}
\usepackage{makecell}
\title{Analytical Delta-V Approximation for Nonlinear Programming of Multi-target Rendezvous and Flyby Trajectories}

\author{An-yi. Huang \footnote{Asscoiate professor, hay04@foxmail.com} and Heng-nian. Li \footnote{Professor, henry\_xscc@mail.xjtu.edu.cn.}}
\affil{State Key Laboratory of Astronautic Dynamics, Xi'an, 710043, China}
\author{Ya-zhong. Luo \footnote{Professor, College of Aerospace Science and Engineering and Hunan Key Laboratory of Intelligent Planning and Simulation for Aerospace Missions, luoyz@nudt.edu.cn.}}
\affil{National University of Defense Technology, Changsha, 410073, China}
\begin{document}
\maketitle
\section{Introduction}
To maximize the benefits of in-orbit service tasks or asteroid exploration missions, multi-target rendezvous or flyby trajectory is usually the first option in the preliminary design and analysis \cite{1,2,3,4}. Trajectory optimization of such missions needs to determine the optimal sequence and visit epochs of each target to minimize the fuel or time, which is a mixed-encoding optimization problem that has been widely studied in \cite{izzo2015, federici2021, huang2022, zhang2023, huang2023, zhang2023b, bellome2024}. \\
When the number of targets is significant, multi-spacecraft missions are more convenient than a single spacecraft to complete more complicated tasks (i.e., in-orbit repairing for a maga constellation or mining of large asteroid clusters). The optimization of such a multi-spacecraft mission requires selecting targets from a large-scale candidate set and assigning them to multiple spacecraft. Similar problems have also been raised by previous Global Trajectory Optimization Competitions (GTOCs) \cite{Izzo2017, Petropoulos2019, Shen2023, Baoyin2023} to promote the development of optimization techniques. Most existing methods focused on the design of global search algorithms with different evolutionary operators \cite{izzo2015, federici2021, huang2022, zhang2023}. Moreover, the cost evaluation of a single-spacecraft sequence also greatly influences efficiency because it must be repeated many times when searching for the optimal target selection and grouping strategy. Several studies have been conducted on the $\Delta v$ approximation for transfers between low-Earth-orbits \cite{roscoe2014, huang2020} and deep space orbits \cite{jifuku2011, guo2023, zhu2018}), which are practical to improve the calculation efficiency. Recently, nonlinear programming models with analytical $\Delta v$ approximations and gradients \cite{huang2023, zhang2023b, bellome2024} are proposed for rapid cost evaluation of a fixed-order rendezvous or flyby sequence, which significantly improve the performance of their global searching algorithms. \\
Therefore, this Note conducts a follow-up study of \cite{zhang2023b}, proposing an analytical $\Delta v$ approximation of short-time transfers based on the linear relative motion and a gradient-based nonlinear programming model of multi-target rendezvous and flyby trajectories. In previous studies, the Lambert's solution is commonly used to evaluate $\Delta v$ of short-duration transfers \cite{federici2021, zhang2023b}. In this Note, to avoid the iteration process for obtaining the Lambert's solution and its gradient, the linear relative motion equations \cite{Tschauner1964, Carter1998, gim2003, dang2018, chernick2018} are applied to form an analytical two-point boundary value model for the near-circular orbit rendezvous problems. Although the relative motion equations are usually applicable when the two orbits are close enough, and the position and velocity errors would become more significant as the orbital differences increase, the errors of the velocity increments were proved acceptable in our simulations. Moreover, the analytical formula facilitates the calculation of the gradients to the start epoch and flight time, which are used to establish a nonlinear programming model for sequence optimization that gradient-based algorithms \cite{gill2005snopt, johnson2021nlopt} can easily solve. Simulation results demonstrated that the analytical $\Delta v$ approximation requires much less calculation than the Lambert's solution, and the proposed gradient-based nonlinear programming algorithms can obtain similar results in less time than previous methods.\\
The rest of this Note is organized as follows. Section II reviews the relative motion and impulsive control around near-circular orbits expressed by the differential orbital elements. Section III proposes the analytical expression for $\Delta v$ of time-fixed orbital transfer and its gradients. Then, Section IV presents a gradient-based non-linear programming algorithm for multi-target rendezvous (or flyby) problems. Section V validates the precision and efficiency of $\Delta v$ calculation and then tests the validity of its application to sequence optimization. Finally, Section VI concludes. 
\section{Relative Motion and Control via Differential Orbital Elements}
The relative motion around a near-circular orbit with near-zero inclination can be expressed by the differential orbital elements proposed in \cite{gim2003}, which are given by
\begin{equation}
\begin{array}{l}
x = \Delta u{a_0} + \frac{{ - 3\Delta a}}{{2{a_0}}}{n_0}t + 2{a_0}\Delta e\sin (u - {u_e})\\
y = \Delta a - {a_0}\Delta e\cos (u - {u_e})\\
z = \Delta i\sin (u - {u_\Omega })
\end{array},
\end{equation}
where $[x, y, z]$ is the relative position in the Vehicle Velocity, Local Horizontal (VVLH) reference frame, $u$ is the argument of latitude (which is approximately calculated as $u=\Omega+\omega+M$ for near-circular orbits), $a_0$ is the reference semimajor axis, $n_0$ is the reference orbital angular velocity (which are set equal to the target orbit), $u_e = \arctan(\Delta e_y/\Delta e_x)$ and $u_o = \arctan(\Delta i_y/\Delta i_x)$ are the arguments of relative perigee and ascending nodes, respectively, and $[\Delta a,\Delta e_x,\Delta e_y,\Delta i_x,\Delta i_y,\Delta l]$ are the differential orbital elements between the chaser and target defined as
\begin{equation}
\begin{array}{l}
\Delta a = a - {a_t}\\
\Delta {e_y} = e\sin (\Omega  + \omega ) - {e_t}\sin ({\Omega _t} + {\omega _t})\\
\Delta {e_x} = e\cos (\Omega  + \omega ) - {e_t}\cos ({\Omega _t} + {\omega _t})\\
\Delta l = \Omega  + \omega  + M - ({\Omega _t} + {\omega _t} + {M_t})\\
\Delta {i_y} = i\sin (\Omega ) - {i_t}\sin ({\Omega _t})\\
\Delta {i_x} = i\cos (\Omega ) - {i_t}\cos ({\Omega _t})
\end{array},
\end{equation}
where $[a, e, i, \Omega, \omega, m]$ and $[a_t, e_t, i_t, \Omega_t, \omega_t, m_t]$ are the classical orbital elements of the initial and target orbits, respectively. Eq. (1) indicates that the relative motion is treated as a combination of three components: the elliptical motion around the target (caused by the difference in eccentricity), tangential drift motion caused by the difference in semimajor axis, and normal motion caused by the difference in inclination.\\
When an impulse is delivered to the spacecraft, the relative position in Eq. (1) remains unchanged. Then, letting $\delta u = \Delta {u^ + } - \Delta {u^ - },\delta a = \Delta {a^ + } - \Delta {a^ - },\delta {e_x} = \Delta e_x^ +  - \Delta e_x^ - ,\delta {e_y} = \Delta e_y^ +  - \Delta e_y^ - ,\delta {i_x} = \Delta i_x^ +  - \Delta i_x^ - ,$ and $\delta {i_y} = \Delta i_y^ +  - \Delta i_y^ - $ denote the changes in the differential orbital elements (where the superscripts '-' and '+' denote the orbital elements before and after the impulse, respectively), one can obtain:
\begin{equation}
\begin{array}{l}
\delta u + 2(\sin {u_0}\delta e_x^{} - \cos {u_0}\delta e_y^{}) = 0\\
\delta a - {a_0}(\cos {u_0}\delta e_x^{} + \sin {u_0}\delta e_y^{}) = 0\\
\sin {u_0}\delta {i_x} - \cos {u_0}\delta i_y^{} = 0
\end{array}.
\end{equation}
Based on the analytical formula for impulsive control in \cite{chernick2018}, the changes in orbital differences are
\begin{equation}
\begin{array}{l}
\delta u = \frac{{2\Delta {v_r}{a_0}}}{{{V_0}}}\\
\delta a = \frac{{2\Delta {v_t}{a_0}}}{{{V_0}}}\\
\delta e_x^{} = \frac{{2\Delta {v_t}}}{{{V_0}}}\cos {u_0} + \frac{{\Delta {v_r}}}{{{V_0}}}\sin {u_0}\\
\delta e_y^{} = \frac{{2\Delta {v_t}}}{{{V_0}}}\sin {u_0} - \frac{{\Delta {v_r}}}{{{V_0}}}\cos {u_0}\\
\delta {i_x} = \frac{{\Delta {v_n}}}{{{V_0}}}\cos {u_0}\\
\delta {i_y} = \frac{{\Delta {v_n}}}{{{V_0}}}\sin {u_0}
\end{array}.
\end{equation}
where $\Delta {v_r}$, $\Delta {v_t}$, and $\Delta {v_n}$ are the three components of the impulse in the VVLH reference frame. According to Eqs. (3) and (4), $\Delta {v_r}$, $\Delta {v_t}$, and $\Delta {v_n}$ corresponding to arbitrary changes in differential orbital elements are expressed by
\begin{equation}
\begin{array}{l}
\Delta {v_r} = \frac{{{V_0}\delta u}}{2}\\
\Delta {v_t} = \frac{{{V_0}\delta a}}{{2{a_0}}}\\
\Delta {v_n} = {V_0}\sqrt {\delta {i_x^2} + \delta {i_y^2}} 
\end{array}.
\end{equation}
Therefore, the analytical motion and impulsive control of a spacecraft in a near-circular orbit can be described by Eqs. (1)-(5). Notably, when the orbits are inclined and the inclinations are close, Eqs. (1)-(5) are still applicable after a rotation of the orbital plane to the equator.
\section{Approximate $\Delta v$ for Orbit Transfer and Its Gradients}
\subsection{Approximate $\Delta v$ via Relative Two-Point Boundary Value Problem}
In this Note, Eqs. (1)-(5) were used to obtain an approximate solution for the two-point boundary value problem with fixed initial and target positions and a fixed transfer duration. Letting $[x_0,y_0,z_0]$ denote the initial position of the spacecraft relative to the target orbit, $\Delta t$ denote the transfer duration, and ${[\Delta a,\Delta e_x,\Delta e_y,\Delta i_x,\Delta i_y,\Delta u]}$ denote the orbital differences of the transfer orbit that connects the initial position to the target orbit results in the derivation of
\begin{equation}
\begin{array}{l}
{x_0} = \Delta u{a_0} + 2{a_0}(\sin {u_0}\Delta {e_x} - \cos {u_0}\Delta {e_y})\\
{y_0} = \Delta a - {a_0}(\cos {u_0}\Delta {e_x} + \sin {u_0}\Delta {e_y})\\
{z_0} = \sin {u_0}\Delta {i_x} - \cos {u_0}\Delta {i_y}\\
{x_f} = \Delta l{a_0} + \frac{{ - 3\Delta a}}{{2{a_0}}}{n_0}\Delta t + 2{a_0}(\sin ({u_0} + {n_0}\Delta t)\Delta {e_x} - \cos ({u_0} + {n_0}\Delta t)\Delta {e_y})\\
{y_f} = \Delta a - {a_0}(\cos ({u_0} + {n_0}\Delta t)\Delta {e_x} + \sin ({u_0} + {n_0}\Delta t)\Delta {e_y})\\
{z_f} = \sin {u_f}\Delta {i_x} - \cos {u_f}\Delta {i_y}
\end{array},
\end{equation}
where $[x_f,y_f,z_f]$ = [0, 0, 0] constraints the spacecraft's position to be the same as that of the target, and $\Delta t$ is the flight time. In the remainder of this Note, we define $\chi$ = $[A=\Delta u, B=\Delta a/{a_0}, C=\Delta e\sin(u-u_e), D=\Delta e\cos(u-u_e), E=\Delta i_x, F=\Delta i_y$] for convenience. Eq. (6) is a linear system of equations with six variables and can be rewritten as
\begin{equation}
\left[ {\begin{array}{*{20}{c}}
1&0&2&0&0&0\\
0&1&0&{ - 1}&0&0\\
0&0&0&0&{\sin ({u_0})}&{\cos ({u_0})}\\
1&{ - 1.5{n_0}\Delta t}&{2\cos ({n_0}\Delta t)}&{2\sin ({n_0}\Delta t)}&0&0\\
0&1&{\sin ({n_0}\Delta t)}&{ - \cos ({n_0}\Delta t)}&0&0\\
0&0&0&0&{\sin ({u_0} + {n_0}\Delta t)}&{\cos ({u_0} + {n_0}\Delta t)}
\end{array}} \right]\left[ {\begin{array}{*{20}{c}}
A\\
B\\
C\\
D\\
E\\
F
\end{array}} \right] = \left[ {\begin{array}{*{20}{c}}
{{x_0}/{a_0}}\\
{{y_0}/{a_0}}\\
{{z_0}/{a_0}}\\
0\\
0\\
0
\end{array}} \right].
\end{equation}
Thus, we can obtain:
\begin{equation}
\begin{array}{l}
D = \frac{1}{{{a_0}}}\frac{{( - {x_0} + 1.5{n_0}\Delta t{y_0})\sin ({n_0}\Delta t) - ({y_f} - {y_0})(2\cos ({n_0}\Delta t) - 2)}}{{( - 1.5{n_0}\Delta t + 2\sin ({n_0}\Delta t))\sin ({n_0}\Delta t) + 2{{(1 - \cos ({n_0}\Delta t))}^2}}}\\
C = \frac{{( - {y_0}/{a_0}) - (1 - \cos ({n_0}\Delta t))D}}{{\sin ({n_0}\Delta t)}}\\
A = \frac{{{x_0}}}{{{a_0}}} - 2C\\
B = \frac{{{y_0}}}{{{a_0}}} + D\\
E = \frac{{{z_0}\cos ({u_0} + {n_0}\Delta t)}}{{ -{a_0} \sin ({n_0}\Delta t)}}\\
F = \frac{{{z_0}\sin ({u_0} + {n_0}\Delta t)}}{{ -{a_0} \sin ({n_0}\Delta t)}}
\end{array}.
\end{equation}
As the Lambert's problem, two impulses in the VVLH frame are required to complete the orbital rendezvous. The first pulse, $\Delta v_{in}$ at the departure epoch, changes the initial orbit to the transfer orbit. The second pulse, $\Delta v_{out}$ after $\Delta t$, changes the orbital differences to zero (the target is $\chi$ = $[0, 0, 0, 0, 0, 0]$). The total velocity increment $\Delta v$ is calculated as
\begin{equation}
\begin{array}{l}
\Delta v = \Delta {v^{in}} + \Delta {v^{out}}\\
\Delta {v^{in}} = {V_0}\sqrt {{{(\frac{{A - A_0}}{2})}^2} + {{(\frac{{B - B_0}}{2})}^2} + {{(E - E_0)}^2} + {{(F - F_0}^2)}} \\
\Delta {v^{out}} = {V_0}\sqrt {{{(\frac{{A - 1.5{n_0}\Delta tB}}{2})}^2} + {{(\frac{B}{2})}^2} + ({E^2} + {F^2})} 
\end{array},
\end{equation}
where $\chi_0$ = $[A_0, B_0, C_0, D_0, E_0, F_0]$ is the initial orbital difference and $- 1.5{n_0}\Delta t B$ in $\Delta v_{out}$ denotes the drift of the relative argument of latitude. 
\subsection{Gradient of $\Delta v$ to the Start Time}
In Eq. (9), $A_0$ of the initial orbit and $\chi$ of the transfer orbit are related to the start time. The gradient of $\Delta v$ to $t_0$ is 
\begin{equation}
\frac{{\partial \Delta v}}{{\partial {t_0}}} = \frac{{\partial \Delta {v^{in}}}}{{\partial {\bf{\chi }}}}\frac{{\partial {\bf{\chi }}}}{{\partial {t_0}}} + \frac{{\partial \Delta {v^{in}}}}{{\partial {A_0}}}\frac{{\partial {A_0}}}{{\partial {t_0}}} + \frac{{\partial \Delta {v^{out}}}}{{\partial {\bf{\chi }}}}\frac{{\partial {\bf{\chi }}}}{{\partial {t_0}}},
\end{equation}
where $\frac{{\partial \Delta {v^{in}}}}{{\partial {A_0}}}\frac{{\partial {A_0}}}{{\partial {t_0}}}$ is the component relative to the initial orbit difference (only $A$ is expressed relative to the time), and $\frac{{\partial \Delta {v^{in}}}}{{\partial {\bf{\chi }}}}$ and $\frac{{\partial \Delta {v^{out}}}}{{\partial {\bf{\chi }}}}$ are expressed relative to the transfer orbit as
\begin{equation}
\begin{array}{l}
\frac{{\partial \Delta {v^{in}}}}{{\partial {A_0}}}\frac{{\partial {A_0}}}{{\partial {t_0}}} = (\frac{{A - {A_0}}}{{4\Delta {v^{in}}}})( - 1.5{n_0}{B_0})\\
\frac{{\partial \Delta {v^{in}}}}{{\partial {\bf{\chi }}}} = {\left[ {\begin{array}{*{20}{c}}
{\frac{{A - \Delta {l_0}}}{{4\Delta {v^{in}}}}}&{\frac{{B - {B_0}}}{{4\Delta {v^{in}}}}}&0&0&{\frac{{E - {E_0}}}{{\Delta {v^{in}}}}}&{\frac{{F - {F_0}}}{{\Delta {v^{in}}}}}
\end{array}} \right]^{\rm{T}}}\\
\frac{{\partial \Delta {v^{out}}}}{{\partial {\bf{\chi }}}} = {\left[ {\begin{array}{*{20}{c}}
{\frac{{A - 1.5{n_0}\Delta tB}}{{4\Delta {v^{out}}}}}&{\frac{{B - 1.5{n_0}\Delta t(A - 1.5{n_0}\Delta tB)}}{{4\Delta {v^{out}}}}}&0&0&{\frac{E}{{\Delta {v^{out}}}}}&{\frac{F}{{\Delta {v^{out}}}}}
\end{array}} \right]^{\rm{T}}}
\end{array}.
\end{equation}
Because Eq. (9) does not include $C$ and $D$ ($\frac{{\partial \Delta {v^{in}}}}{{\partial C}} = \frac{{\partial \Delta {v^{in}}}}{{\partial D}} = \frac{{\partial \Delta {v^{out}}}}{{\partial C}} = \frac{{\partial \Delta {v^{out}}}}{{\partial D}} = 0$), the corresponding terms are removed from $\frac{{\partial \Delta {v^{in}}}}{{\partial {\bf{\chi }}}}$, $\frac{{\partial \Delta {v^{out}}}}{{\partial {\bf{\chi }}}}$ and $\frac{{\partial {\bf{\chi }}}}{{\partial {t_0}}}$. Therefore, $\frac{{\partial {\bf{\chi }}}}{{\partial {t_0}}}$ can be expressed as
\begin{equation}
\frac{{\partial {\bf{\chi }}}}{{\partial {t_0}}} = \frac{{\partial {\bf{\chi }}}}{{\partial {{\bf{r}}_0}}}\frac{{\partial {{\bf{r}}_0}}}{{\partial {t_0}}} = \left[ {\begin{array}{*{20}{c}}
{\frac{{\partial A}}{{\partial {x_0}}}}&{\frac{{\partial A}}{{\partial {y_0}}}}&{\frac{{\partial A}}{{\partial {z_0}}}}\\
{\frac{{\partial B}}{{\partial {x_0}}}}&{\frac{{\partial B}}{{\partial {y_0}}}}&{\frac{{\partial B}}{{\partial {z_0}}}}\\
{\frac{{\partial E}}{{\partial {x_0}}}}&{\frac{{\partial E}}{{\partial {y_0}}}}&{\frac{{\partial E}}{{\partial {z_0}}}}\\
{\frac{{\partial F}}{{\partial {x_0}}}}&{\frac{{\partial F}}{{\partial {y_0}}}}&{\frac{{\partial F}}{{\partial {z_0}}}}
\end{array}} \right]\left[ {\begin{array}{*{20}{c}}
{\frac{{\partial {x_0}}}{{\partial {t_0}}}}\\
{\frac{{\partial {y_0}}}{{\partial {t_0}}}}\\
{\frac{{\partial {z_0}}}{{\partial {t_0}}}}
\end{array}} \right].
\end{equation}
The nonzero components of $\frac{{\partial {\bf{\chi }}}}{{\partial {{\bf{r}}_0}}}$ and $\frac{{\partial {{\bf{r}}_0}}}{{\partial {t_0}}}$ are calculated via 
\begin{equation}
\begin{array}{l}
\frac{{\partial {x_0}}}{{\partial {t_0}}} = \frac{{ - 3\Delta a}}{{2{a_0}}}{n_0} + 2{a_0}{n_0}D\\
\frac{{\partial {y_0}}}{{\partial {t_0}}} = {a_0}{n_0}C\\
\frac{{\partial {z_0}}}{{\partial {t_0}}} = {n_0}F\\
\frac{{\partial D}}{{\partial {x_0}}} = \frac{1}{{{a_0}}}\frac{{ - \sin ({n_0}\Delta t)}}{{( - 1.5{n_0}\Delta t + 2\sin ({n_0}\Delta t))\sin ({n_0}\Delta t) + 2{{(1 - \cos ({n_0}\Delta t))}^2}}}\\
\frac{{\partial D}}{{\partial {y_0}}} = \frac{1}{{{a_0}}}\frac{{1.5{n_0}\Delta t\sin ({n_0}\Delta t) + (2\cos ({n_0}\Delta t) - 2)}}{{( - 1.5{n_0}\Delta t + 2\sin ({n_0}\Delta t))\sin ({n_0}\Delta t) + 2{{(1 - \cos ({n_0}\Delta t))}^2}}}\\
\frac{{\partial C}}{{\partial {x_0}}} = \frac{{\cos ({n_0}\Delta t)}}{{\sin ({n_0}\Delta t)}}\frac{{\partial D}}{{\partial {x_0}}}\\
\frac{{\partial C}}{{\partial {y_0}}} = \frac{{ - \frac{1}{{{a_0}}} + \cos ({n_0}\Delta t)\frac{{\partial D}}{{\partial {y_0}}}}}{{\sin ({n_0}\Delta t)}}\\
\frac{{\partial A}}{{\partial {x_0}}} = \frac{1}{{{a_0}}} - 2\frac{{\partial C}}{{\partial {x_0}}}\\
\frac{{\partial A}}{{\partial {y_0}}} = 2\frac{{\partial C}}{{\partial {y_0}}}\\
\frac{{\partial B}}{{\partial {x_0}}} = \frac{{\partial D}}{{\partial {x_0}}}\\
\frac{{\partial B}}{{\partial {y_0}}} = \frac{1}{{{a_0}}} + \frac{{\partial D}}{{\partial {y_0}}}\\
\frac{{\partial E}}{{\partial {z_0}}} = \frac{{\cos ({u_0} + {n_0}\Delta t)}}{{ -{a_0} \sin ({n_0}\Delta t)}}\\
\frac{{\partial F}}{{\partial {z_0}}} = \frac{{\sin ({u_0} + {n_0}\Delta t)}}{{ -{a_0} \sin ({n_0}\Delta t)}}
\end{array}.
\end{equation}
\subsection{Gradient of $\Delta v$ to the Transfer Duration}
The gradient of $\Delta v$ with respect to the transfer duration $\Delta t$ is
\begin{equation}
\frac{{\partial \Delta v}}{{\partial \Delta t}} = \frac{{\partial \Delta {v^{in}}}}{{\partial \Delta t}} + \frac{{\partial \Delta {v^{out}}}}{{\partial \Delta t}} = \frac{{\partial \Delta {v^{in}}}}{{\partial {\bf{\chi }}}}\frac{{\partial {\bf{\chi }}}}{{\partial \Delta t}} + \frac{{\partial \Delta {v^{out}}}}{{\partial {\bf{\chi }}}}\frac{{\partial {\bf{\chi }}}}{{\partial \Delta t}}
\end{equation}
where $\frac{{\partial \Delta {v^{in}}}}{{\partial {\bf{\chi }}}}$ and $\frac{{\partial \Delta {v^{out}}}}{{\partial {\bf{\chi }}}}$ are the same as in Eq. (11). The $\frac{{\partial {\bf{\chi }}}}{{\partial \Delta t}}$ term is calculated as
\begin{equation}
\begin{array}{l}
\frac{{\partial A}}{{\partial \Delta t}} =  - 2\frac{{\partial C}}{{\partial \Delta t}}\\
\frac{{\partial B}}{{\partial \Delta t}} = \frac{{\partial D}}{{\partial \Delta t}}\\
\frac{{\partial C}}{{\partial \Delta t}} = \frac{{ - {n_0}\sin ({n_0}\Delta t))D}}{{\sin ({n_0}\Delta t)}} + \frac{{\cos ({n_0}\Delta t))}}{{\sin ({n_0}\Delta t)}}\frac{{\partial D}}{{\partial \Delta t}} - \frac{{( - {y_0}/{a_0}) - (1 - \cos ({n_0}\Delta t))D}}{{{{\sin }^2}({n_0}\Delta t)}}{n_0}\cos ({n_0}\Delta t)\\
\frac{{\partial D}}{{\partial \Delta t}} = \frac{1}{{{a_0}}}(\frac{{{n_0}( - {x_0} + 1.5{n_0}\Delta t{y_0})\cos ({n_0}\Delta t) + 1.5{n_0}{y_0}\sin ({n_0}\Delta t) - ({y_0})(2{n_0}\sin ({n_0}\Delta t))}}{{( - 1.5{n_0}\Delta t + 2\sin ({n_0}\Delta t))\sin ({n_0}\Delta t) - 2(1 - {{\cos }^2}({n_0}\Delta t))}}\\
 - \frac{{{a_0}D(( - 1.5{n_0} + 2{n_0}\cos ({n_0}\Delta t))\sin ({n_0}\Delta t) + ( - 1.5{n_0}\Delta t + 2\sin ({n_0}\Delta t)){n_0}\cos ({n_0}\Delta t) + 4(1 - \cos ({n_0}\Delta t)){n_0}\sin ({n_0}\Delta t))}}{{( - 1.5{n_0}\Delta t + 2\sin ({n_0}\Delta t))\sin ({n_0}\Delta t) - 2(1 - {{\cos }^2}({n_0}\Delta t))}})\\
\frac{{\partial E}}{{\partial \Delta t}} = \frac{{{z_0}{n_0}\sin ({u_0} + {n_0}\Delta t)}}{{{a_0}\sin ({n_0}\Delta t)}} + \frac{{{z_0}\cos ({u_0} + {n_0}\Delta t)}}{{{{a_0}{\sin }^2}({n_0}\Delta t)}}{n_0}\cos ({n_0}\Delta t)\\
\frac{{\partial F}}{{\partial \Delta t}} = \frac{{ - {z_0}{n_0}\cos ({u_0} + {n_0}\Delta t)}}{{{a_0}\sin ({n_0}\Delta t)}} + \frac{{{z_0}\sin ({u_0} + {n_0}\Delta t)}}{{{{a_0}{\sin }^2}({n_0}\Delta t)}}{n_0}\cos ({n_0}\Delta t)
\end{array}.
\end{equation}
Thus, the gradients of $\Delta v$ to the start time and flight duration are obtained, and they can be applied to the sequence optimization of multi-target rendezvous or flyby missions.
\section{Gradient-Based Nonlinear Programming of Multi-Target Sequence}
In this section, the multi-target sequence optimization problem of fixed target order is formulated, and the analytical gradients of the objective function and constraints are derived. Subsequently, the solving process of the gradient-based nonlinear programming algorithm is proposed.
\subsection{Problem Formulation}
A similar multi-target rendezvous or flyby problem was in \cite{huang2023, zhang2023b}. The spacecraft is assumed to start from a given orbit and then sequentially rendezvous with (or flyby) each target in a fixed order. The earliest start and latest end times are constrained, while the arrival time of each target is unconstrained. Sequence optimization involves determining the optimal flight time for each transfer that minimizes the total velocity increment, which determines the fuel cost of the mission. \\
\begin{figure}[hbt!]
\centering
\includegraphics[width=0.5\textwidth]{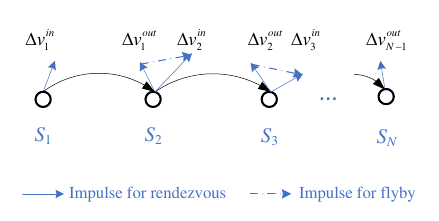}
\caption{Impulsive multi-target mission.}
\label{fig:fig1}
\end{figure}
This Note further assumes that the transfer duration between every two targets is much shorter than their orbital periods and that the double-impulse transfer strategy is optimal. Including impulses in the transfer medium is unnecessary, as illustrated in Fig. \ref{fig:fig1}. Assume that the set of targets is ${S_i}$ (where $i=1,2…N$, $N$ is the number of targets, and the order is fixed). Let $t_0$ and $t_f$ denote the earliest start time and latest end time, respectively. The flight time of the transfer from $S_i$ to $S_{i+1}$ is $\Delta t_i$ ($i=1,1,2…N$), which can be expressed by a group of normalized variables $\bf{x}=\it\{x_1,x_2...x_{N}\}$ via
\begin{equation}
\Delta {t_i} = (\Delta _t^{\max } - \Delta _t^{\min }){x_i} + \Delta _t^{\min },
\end{equation}
where $\Delta _t^{\max }$ and $\Delta _t^{\min }$ are the maximum and minimum flight times, respectively. Note that $\Delta {t_0}$ means the wait time before the spacecraft leaves the first target. Thus, the arrival (and departure) times of each target are
\begin{equation}
{t_i} = {t_0} + \sum\limits_{k = 1}^{i - 1} {\Delta {t_k}} .
\end{equation}
The constraint on the mission time is:
\begin{equation}
{t_0} + \sum\limits_{i = 0}^{N - 1} {\Delta {t_i}}  - {t_f} \le 0.
\end{equation}
For the multi-target rendezvous problem, the objective function minimizes the velocity increment via
\begin{equation}
{J_r} = \sum\limits_{i = 1}^{N - 1} {\Delta {v_i}({S_i},{S_{i + 1}},{t_i},\Delta {t_i})} ,
\end{equation}
where $\Delta {v_i}({S_i},{S_{i + 1}},{t_i},\Delta {t_i})$ is the velocity increment required for the $i^{th}$transfer from $S_i$ to $S_{i+1}$, which is the sum of $\Delta v_{i}^{in}$ and $\Delta v_{i}^{out}$ shown in Fig. 1, and can be evaluated using Eq. (9). In Eq. (9) $\bf{\chi }^{\rm{0}}$ should be replaced by $\bf{\chi}_{\it i}^0$ (the orbital differences between the $i^{th}$ target and the next target).\\
For the multi-target flyby problem, the velocity increment required at $t_i$ is related to both the $i^{th}$ and ${(i-1)}^{th}$ transfers because the spacecraft directly changes its velocity from $v^{out}_{i-1}$ to $v^{in}_{i}$, as illustrated in Fig. 1. Therefore, the objective function minimizes the velocity increment via
\begin{equation}
{J_f} = \sum\limits_{i = 1}^{N - 1} {\Delta {v_i}({{\bf{\chi }}_{i - 1}},{{\bf{\chi }}_i})} ,
\end{equation}
where $\Delta {v_i}({{\bf{\chi }}_{i - 1}},{{\bf{\chi }}_i})$ indicates the velocity increment is related to the relative orbits before and after visiting the $i^{th}$ target ($\Delta v_1$ remains equal to $\Delta v_1^{in}$ in Fig. 1), and is calculated via
\begin{equation}
\begin{array}{l}
\Delta {v_i}({{\bf{\chi }}_{i - 1}},{{\bf{\chi }}_i}) = \left| {\Delta {{\bf{v}}_{i - 1}^{out}} - \Delta {{\bf{v}}_i^{in}}} \right|
 = \left| {{V_0}\left[ {\begin{array}{*{20}{c}}
{\frac{{({A_{i}} - A_i^0) - (0 - {A_{i-1}} + 1.5{n_0}\Delta t{B_{i-1}})}}{2}}\\
{\frac{{({B_{i}} - B_i^0) - (0 - {B_{i-1}})}}{2}}\\
{\sqrt {{(E_{i}-E_i^0)^2} + {(F_{i} - F_i^0)^2}}  - \sqrt {{E_{i-1}^2} + {F_{i-1}^2}} }
\end{array}} \right]} \right|
\end{array}.
\end{equation}
Because the argument of latitude, $u_i$, remains the same before and after the impulse, we get $(E_i - E_i^0)/(F_i -F_i^0 ) = E_{i-1}/F_{i-1} = \tan(u_i)$ and apply it to Eq. (21):
\begin{equation}
\sqrt {{(E_{i}-E_i^0)^2} + {(F_{i} - F_i^0)^2}}  - \sqrt {{E_{i-1}^2} + {F_{i-1}^2}}  = \sqrt {{{({E_{i}} - E_i^0 - {E_{i-1}}))}^2} + {{({F_{i}} - F_i^0 - {F_{i-1}})}^2}} 
\end{equation}
Both optimization problems of orbit rendezvous (Eq. (19)) or flyby (Eq. (20)) can be solved by the traditional evolutionary \cite{huang2022} and nonlinear programming algorithms \cite{huang2023}. In this Note, as a result of the analytical expression derived for the constraints and objective functions, efficient gradient-based optimization algorithms (in particular, the sequential quadratic programming (SQP) algorithm \cite{gill2005snopt, johnson2021nlopt}) can be applied. The analytical gradients are given in the following subsections.
\subsection{Analytical Gradient of Multi-Target Rendezvous Mission}
The gradient of time constraint (Eq. (12)) is
\begin{equation}
\frac{{\partial \sum\limits_{k = 1}^N {\Delta {t_k}} }}{{\partial {x_i}}} = (\Delta _t^{\max } - \Delta _t^{\min }).
\end{equation}
The gradient of $J_r$ is
\begin{equation}
\frac{{\partial \sum\limits_{k = 1}^{N - 1} {\Delta {v_k}} }}{{\partial {x_i}}} = \sum\limits_{k = 1}^{N - 1} {(\frac{{\partial \Delta {v_k}}}{{\partial {t_{k - 1}}}}\frac{{\partial {t_{k - 1}}}}{{\partial {x_i}}} + \frac{{\partial \Delta {v_k}}}{{\partial \Delta {t_k}}}\frac{{\partial \Delta {t_k}}}{{\partial {x_i}}})} 
\end{equation}
where $\frac{{\partial \Delta {v_k}}}{{\partial {t_{k - 1}}}}$ and $\frac{{\partial \Delta {v_k}}}{{\partial \Delta {t_k}}}$ are calculated by Eqs. (10) and (14). $\frac{{\partial {t_{k - 1}}}}{{\partial {x_i}}}$ and $\frac{{\partial \Delta {t_k}}}{{\partial {x_i}}}$ are calculated via
\begin{equation}
\begin{array}{l}
\frac{{\partial {t_{k - 1}}}}{{\partial {x_i}}} = \left\{ \begin{array}{l}
\sum\limits_{k = 1}^{i - 1} {(\Delta _t^{\max } - \Delta _t^{\min }),k - 1 \ge i} \\
0,k - 1 < i
\end{array} \right.\\
\frac{{\partial \Delta {t_k}}}{{\partial {x_i}}} = \left\{ \begin{array}{l}
(\Delta _t^{\max } - \Delta _t^{\min }),k = i\\
0,k \ne i
\end{array} \right.
\end{array}.
\end{equation}
\subsection{Analytical Gradient of Multi-Target Flyby Mission}
 The time constraint gradient is given by Eq. (18), and the gradient of $\Delta v$ in Eq. (15) is calculated according to
\begin{equation}
\frac{{\partial \Delta {v_k}}}{{\partial {x_i}}} = \frac{{\partial \Delta {v_k}}}{{\partial {{\bf{\chi }}_{i - 1}}}} \cdot {(\frac{{\partial {{\bf{\chi }}_{i - 1}}}}{{\partial {t_{i - 1}}}}\frac{{\partial {t_{i - 1}}}}{{\partial {x_i}}} + \frac{{\partial {{\bf{\chi }}_{i - 1}}}}{{\partial \Delta {t_{i - 1}}}}\frac{{\partial \Delta {t_{i - 1}}}}{{\partial {x_i}}})^{\rm{T}}} + \frac{{\partial \Delta {v_k}}}{{\partial {{\bf{\chi }}_i}}} \cdot {(\frac{{\partial {{\bf{\chi }}_i}}}{{\partial {t_{i + 1}}}}\frac{{\partial {t_{i + 1}}}}{{\partial {x_i}}} + \frac{{\partial {{\bf{\chi }}_i}}}{{\partial \Delta {t_i}}}\frac{{\partial \Delta {t_i}}}{{\partial {x_i}}})^{\rm{T}}} + \frac{{\partial \Delta {v_k}}}{{\partial {\bf{\chi }}_i^0}} \cdot {(\frac{{\partial {\bf{\chi }}_i^0}}{{\partial {t_i}}}\frac{{\partial {t_i}}}{{\partial {x_i}}})^{\rm{T}}},
\end{equation}
where $\frac{{\partial \Delta {v_k}}}{{\partial {{\bf{\chi }}_{i - 1}}}}$, $\frac{{\partial \Delta {v_k}}}{{\partial {{\bf{\chi }}_i}}}$, and $\frac{{\partial \Delta {v_k}}}{{\partial {\bf{\chi }}_i^0}}$ are given by
\begin{equation}
\begin{array}{l}
\frac{{\partial \Delta {v_i}}}{{\partial {A_i}}} = \frac{{{V_0}^2}}{{\Delta {v_i}}}(\frac{{{A_i} - 1.5{n_0}\Delta t{B_i} - {A_{i + 1}}}}{4})\\
\frac{{\partial \Delta {v_i}}}{{\partial {B_i}}} = \frac{{{V_0}^2}}{{\Delta {v_i}}}( - 1.5{n_0}\Delta t(\frac{{{A_i} - 1.5{n_0}\Delta t{B_i} - {A_{i + 1}}}}{4}) + (\frac{{{B_i} - {B_{i + 1}}}}{4}))\\
\frac{{\partial \Delta {v_i}}}{{\partial {E_i}}} = \frac{{{V_0}^2}}{{\Delta {v_i}}}({E_i} - {E_{i + 1}})\\
\frac{{\partial \Delta {v_i}}}{{\partial {F_i}}} = \frac{{{V_0}^2}}{{\Delta {v_i}}}({F_i} - {F_{i + 1}})\\
\frac{{\partial \Delta {v_i}}}{{\partial {A_{i + 1}}}} = \frac{{{V_0}^2}}{{\Delta {v_i}}}(\frac{{{A_i} - 1.5{n_0}\Delta t{B_i} - {A_{i + 1}}}}{4})\\
\frac{{\partial \Delta {v_i}}}{{\partial {B_{i + 1}}}} = \frac{{{V_0}^2}}{{\Delta {v_i}}}(\frac{{{B_i} - {B_{i + 1}}}}{4})\\
\frac{{\partial \Delta {v_i}}}{{\partial {E_{i + 1}}}} = \frac{{{V_0}^2}}{{\Delta {v_i}}}({E_i} - {E_{i + 1}})\\
\frac{{\partial \Delta {v_i}}}{{\partial {F_{i + 1}}}} = \frac{{{V_0}^2}}{{\Delta {v_i}}}({F_i} - {F_{i + 1}})\\
\frac{{\partial \Delta {v_i}}}{{\partial A_i^0}} =  - \frac{{{V_0}^2}}{{\Delta {v_i}}}(\frac{{{A_i} - 1.5{n_0}\Delta t{B_i} + {A_{i + 1}} - A_i^0}}{4})\\
\frac{{\partial \Delta {v_i}}}{{\partial B_i^0}} =  - \frac{{{V_0}^2}}{{\Delta {v_i}}}(\frac{{{B_i} + {B_{i + 1}} - B_i^0}}{4})\\
\frac{{\partial \Delta {v_i}}}{{\partial E_i^0}} =  - \frac{{{V_0}^2}}{{\Delta {v_i}}}({E_i} + {E_{i + 1}} - E_i^0)\\
\frac{{\partial \Delta {v_i}}}{{\partial F_i^0}} =  - \frac{{{V_0}^2}}{{\Delta {v_i}}}({F_i} + {F_{i + 1}} - F_i^0)
\end{array}.
\end{equation}
The calculations of $\frac{{\partial {{\bf{\chi }}_{i - 1}}}}{{\partial {t_{i - 1}}}}$, $\frac{{\partial {{\bf{\chi }}_i}}}{{\partial {t_{i + 1}}}}$, $\frac{{\partial {{\bf{\chi }}_{i - 1}}}}{{\partial \Delta {t_{i - 1}}}}$, and $\frac{{\partial {{\bf{\chi }}_i}}}{{\partial \Delta {t_i}}}$ can be performed using Eqs. (13) and (16), and $\frac{{\partial \Delta {v_k}}}{{\partial {\bf{\chi }}_i^0}}$ and $\frac{{\partial \Delta {v_k}}}{{\partial {\bf{\chi }}_i^0}}$ (only $\frac{{\partial A_i^0}}{{\partial {t_i}}}$ is nonzero) can be calculated by Eq. (12).
\subsection{Solving Process}
As the gradients of the objective function and constraints are obtained, the SQP algorithm can be applied to solve such multi-target missions. In this Note, NLOPT \cite{johnson2021nlopt}, an open-source library including a SQP solver, is used. The models can be solved directly by defining the objective function, constraints, and their gradients. More details on the SQP can be found in \cite{johnson2021nlopt}. 
\section{Simulation and Validation}
This section analyzes the orbit transfers between asteroids in the main belt to test the proposed $\Delta v$ evaluation method and the multi-target sequence optimization algorithm. First, the precision of $\Delta v$ approximation and its gradients are validated. Second, a multi-asteroid mission from the champion’s solution \cite{GTOC12JPLSol} of the 12th Global Trajectory Optimization Competition (GTOC12) \cite{Baoyin2023} is used to test the gradient-based nonlinear programming algorithm. 
\subsection{Validation of $\Delta v$ Evaluation}
A dataset of 600000 transfers between different asteroids (randomly selected in the dataset given by \cite{Baoyin2023}) is generated, and the $\Delta v$ is calculated by the Lambert's solution and the proposed approximate method to make the comparison (all of the $\Delta v$ are less than 10,000 m/s). The mean relative errors corresponding to different transfer durations and ranges of eccentricity and inclination are summarized in Table 1, which indicates that the approximation model via the relative motion is acceptable for sequence optimization even though the eccentricity and inclination are significant (0.1$\sim$0.2). A multi-target mission typically includes a target cluster of close orbital elements. Therefore, higher differences in eccentricity or inclination are infrequent and are not considered in this Note. 
\begin{table}[hbt!]
\caption{\label{tab:table1} Relative error of $\Delta v$ approximation}
\centering
%\resizebox{\linewidth}{!}{
\begin{tabular}{lcccccc}
\hline
\hline
& Flight time (d)&\makecell[c]{ Mean relative error \\($\Delta e < 0.2$, $\Delta i < 0.2$)}	& \makecell[c]{Mean relative error \\($\Delta e < 0.1$, $\Delta i < 0.1$)}\\\hline
&60$\sim$300&	4.52\%	&3.83\%\\
&60	&10.67\%	&7.67\%\\
&120	&6.29\%	&4.57\%\\
&210	&4.38\%	&3.65\%\\
&300	&4.06\%	&3.56\%\\
\hline
\hline
\end{tabular}
%}
\end{table}

\begin{figure*}[!t]
\centering
\subfloat[$\Delta e < 0.2$, $\Delta i < 0.2$]{\includegraphics[width=0.4\textwidth]{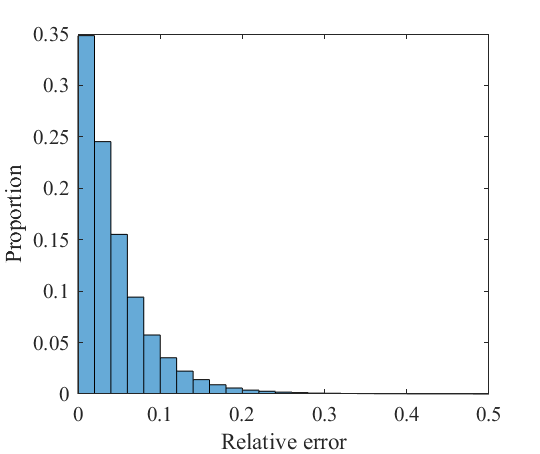}}
\label{subfig2a}
\hfil
\subfloat[$\Delta e < 0.1$, $\Delta i < 0.1$]{\includegraphics[width=0.4\textwidth]{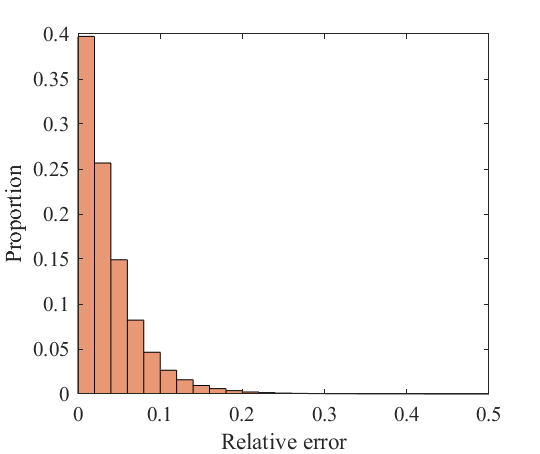}}
\label{subfig2b}
\caption{Diagram of the relative error.}
\label{fig:fig2}
\end{figure*}
Compared to the Lambert's solution, the approximation method based on geometric relative motion equations is analytical and requires no iteration. The time required for calculation is less than 17\% of that required by the Lambert's solution on a desktop computer ($5.5\times 10^{-7}$ s vs $3.2\times 10^{-6}$ s). The gradient can also be directly obtained during the calculation with little additional calculation. By contrast, the Lambert's solution requires more than 50\% calculation to obtain the gradient \cite{zhang2023b}. \\
To further validate the evaluations of $\Delta v$ and the corresponding gradients, two typical transfers listed in Tables 2 and 3 are analyzed. Results for different departure times and transfer durations are illustrated in Figures 3 and 4, respectively. When the semimajor axis difference is more significant than the differences in eccentricity and inclination, $\Delta u$ primarily determines the velocity increment. When the semimajor axis difference is slight, $\Delta v$ is mainly related to the argument of latitude and changes periodically. Overall, the results obtained using the proposed method and the Lambert's solution are very similar. The analytical and central differential gradients are illustrated in Fig. 5, respectively. These figures demonstrate the calculation's validity. 
\begin{table}[hbt!]
\caption{\label{tab:table2} Initial orbits of Case 1}
\centering
\begin{tabular}{lccccccc}
\hline
\hline
&Asteroid ID	&$a$ (AU)	&$e$	&$i$ (deg)	&$\Omega$ (deg)	&$\omega$(deg)	&$M$ (deg)\\\hline
&4184	&2.781	&0.0681	&4.7	&32.12	&12.38	&122.3105\\
&31302	&2.800	&0.054	&2.67	&13.45	&287.19	&251.2186\\
\hline
\hline
\end{tabular}
\end{table}
\begin{table}[hbt!]
\caption{\label{tab:table3} Initial orbits of Case 2}
\centering
\begin{tabular}{lccccccc}
\hline
\hline
&Asteroid ID	&$a$ (AU)	&$e$	&$i$ (deg)	&$\Omega$ (deg)	&$\omega$(deg)	&$M$ (deg)\\\hline
&31976	&2.677		&0.0451		&2.9		&178.5		&301.33		&233.253\\
&1546	&2.967		&0.0506		&5.42		&148.28		&346.45		&284.595\\
\hline
\hline
\end{tabular}
\end{table}

\begin{figure*}[hbt!]
\centering
\subfloat[Case 1]{\includegraphics[width=0.45\textwidth,height=0.34\textwidth]{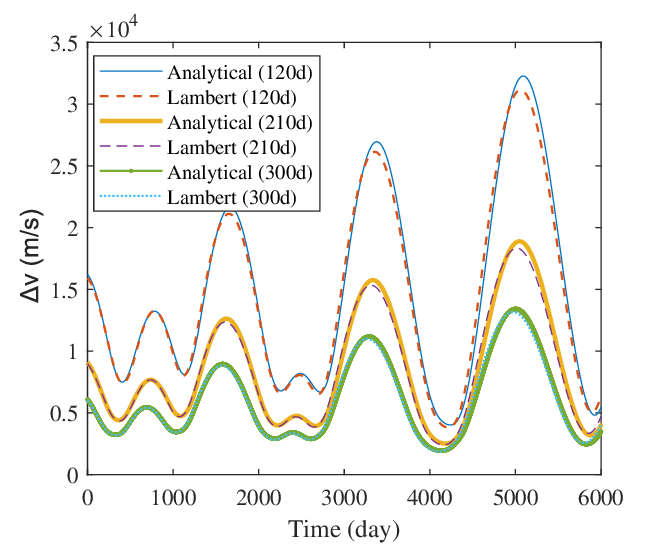}}
\label{subfig3a}
\hfil
\subfloat[Case 2]{\includegraphics[width=0.45\textwidth,height=0.34\textwidth]{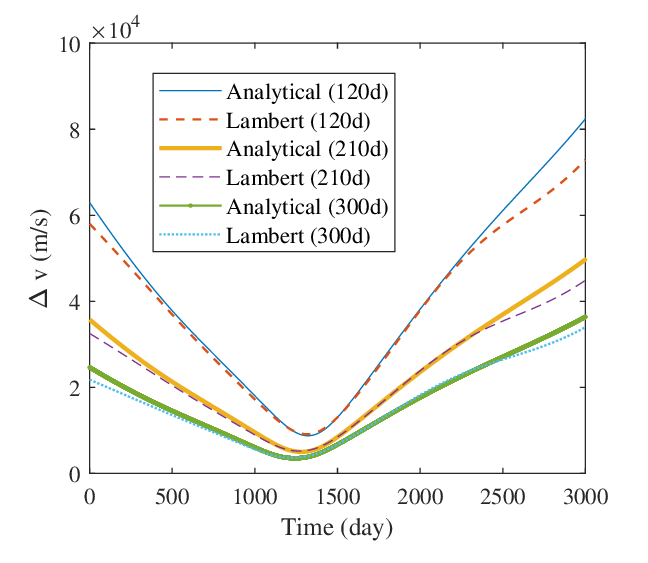}}
\label{subfig3b}
\caption{Validation of $\Delta v$ approximation.}
\label{fig:fig3}
\end{figure*}
\begin{figure*}[hbt!]
\centering
\subfloat[Case 1]{\includegraphics[width=0.45\textwidth,height=0.34\textwidth]{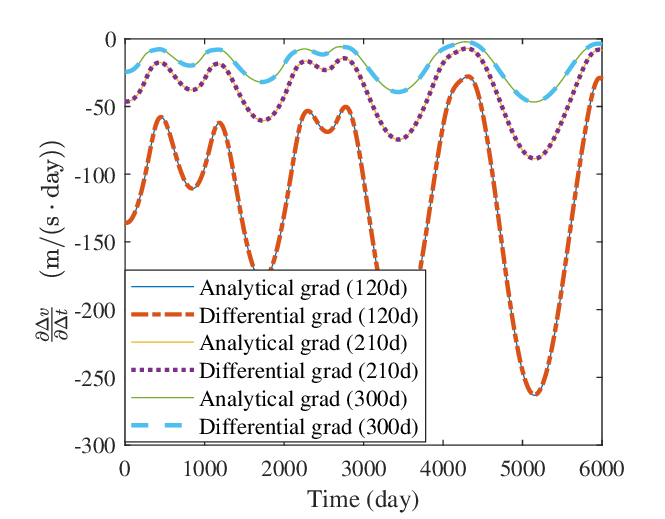}}
\label{subfig4a}
\hfil
\subfloat[Case 2]{\includegraphics[width=0.45\textwidth,height=0.34\textwidth]{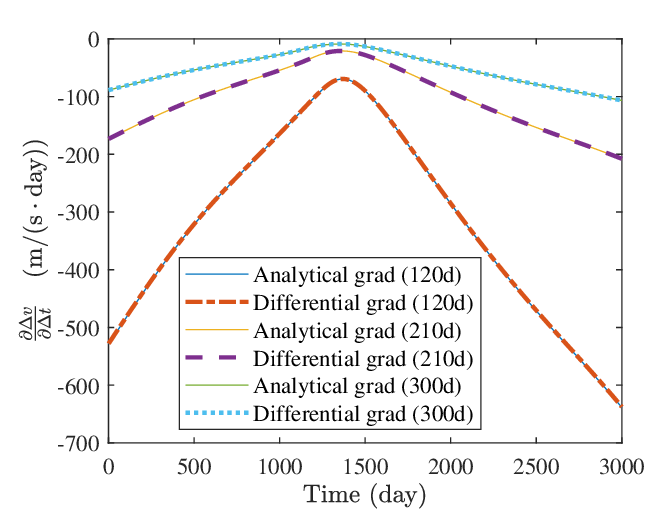}}
\label{subfig4b}
\caption{Validation of $\frac{{\partial \Delta v}}{{\partial \Delta t}}$.}
\label{fig:fig4}
\end{figure*}
\begin{figure*}[hbt!]
\centering
\subfloat[Case 1]{\includegraphics[width=0.45\textwidth,height=0.34\textwidth]{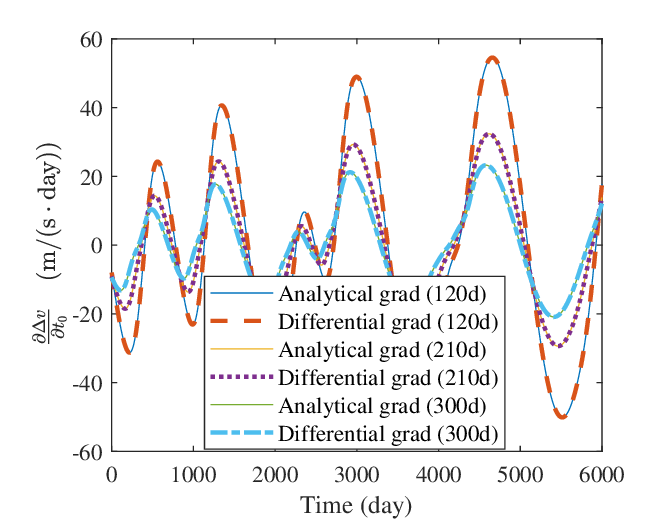}}
\label{subfig5a}
\hfil
\subfloat[Case 2]{\includegraphics[width=0.45\textwidth,height=0.34\textwidth]{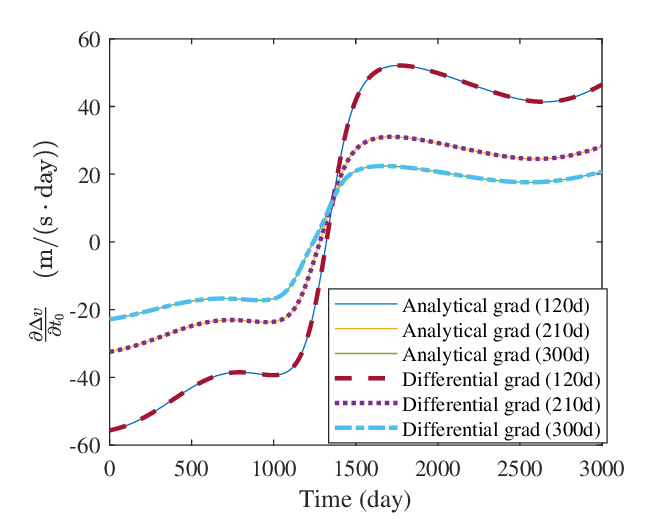}}
\label{subfig5b}
\caption{Validation of $\frac{{\partial \Delta v}}{{\partial t_0}}$.}
\label{fig:fig5}
\end{figure*}

\subsection{Validation of Gradient-Based Nonlinear Programming Algorithm}
As described in \cite{Baoyin2023}, the GTOC12 problem is a multi-spacecraft and multi-target global optimization problem. All the selected targets should be visited twice to complete the mining task. The objective function maximizes the sum of the collected masses (related to the time between every two visits) under the constraint of limited fuel, which increases the complexity of the problem compared to the fuel-optimal problem described in Section V.C. Because global optimization of such multi-spacecraft is not the focus of this Note, we choose the first chain of the champion’s solution \cite{GTOC12JPLSol} to validate the proposed gradient-based sequence optimization method, and only considered the first rendezvous of each target. \\
The nine target orbits are listed in Table 3, and their order is fixed. The spacecraft must start from the first asteroid; the start time should be no earlier than $t_0$ = 546 d, and the arrival time of the last target should not be later than $t_f$ = 2,400 d. The initial values of all flight times are set to $(t_f-t_0)/(N-1)$. 
\begin{table}[hbt!]
\caption{\label{tab:table4} Orbital elements of the targets at $t_0$}
\centering
\begin{tabular}{lccccccc}
\hline
\hline
&Asteroid ID	&$a$ (AU)	&$e$	&$i$ (deg)	&$\Omega$ (deg)	&$\omega$(deg)	&$M$ (deg)\\\hline
&12095	&2.755	&0.0464	&2.46	&128.42	&184.82	&262.151\\
&3506	&2.756	&0.076	&5.24	&18.31	&357.75	&207.11\\
&49192	&2.784	&0.0752	&4.66	&33.17	&6.94	&187.782\\
&33590	&2.789	&0.0444	&4.62	&24.03	&49.86	&159.639\\
&36666	&2.781	&0.0253	&3.86	&350.57	&108.19	&135.107\\
&2154	&2.84&	0.0078	&1.71	&9.69	&141.24	&94.356\\
&33908	&2.788	&0.0772	&4.45	&16.45	&19.91	&198.185\\
&35666	&2.791	&0.0058	&4.75	&22.25	&172.86	&33.1496\\
&4971	&2.857	&0.044	&2.45	&83.03	&164.73	&352.666\\
\hline
\hline
\end{tabular}
\end{table}
The optimal rendezvous times for each target obtained using the gradient-based method are listed in Table 4. The total velocity increment ($J_r$) is 15,351.22 m/s (the result obtained by the Lambert's solution is 15,413.28 m/s), which is 85\% of the solution submitted by JPL \cite{GTOC12JPLSol}. Note that the solution in GTOC12 was asked to use low-thrust propulsion, which requires about 20\% more velocity increment in previous studies. The results obtained by replacing the geometric model with the Lambert's solution (similar to \cite{zhang2023b}) and replacing the gradient-based optimization algorithm with a differential evolution (DE) algorithm (similar to \cite{huang2022}) are also listed in Table 4. It’s proved that the result of the proposed method is consistent with existing methods. The results using different random initial values are illustrated by a box plot in Fig. 6, which indicates the max relative error of local optimal solutions is less than 10\%. \\
A significant advantage of the proposed method is its high calculation efficiency. The histories of the objective function by different optimization methods are illustrated in Figures 7 and 8. The proposed method converges in less than $10^{-4}$ s. In contrast, when the geometric model is replaced with the Lambert's solution, the convergence requires $6.4\times 10^{-4}$ s. Convergence of DE also requires much more objective function evaluations.
\begin{table}[hbt!]
\caption{\label{tab:table5} Optimal solutions of multi-target rendezvous mission}
\centering
\resizebox{\linewidth}{!}{
\begin{tabular}{lccccccccc}
\hline
\hline
\multirow{2}{*}{Asteroid ID} &\multicolumn{2}{c}{Optimal solution by SQP}	&\multicolumn{2}{c}{Optimal solution by DE}&\multicolumn{2}{c}{Optimal solution by SQP (Lambert)}	&\multicolumn{2}{c}{Optimal solution by DE (Lambert)}\\
\cline{2-9}
&Arrive time (d) &$\Delta v$ (m/s) &Arrive time (d) &$\Delta v$ (m/s) &Arrive time (d) &$\Delta v$ (m/s) &Arrive time (d) &$\Delta v$ (m/s)\\\hline
12095	&546	&$\backslash$	&546	&$\backslash$	&546	&$\backslash$	&546	&$\backslash$ \\
3506	&670.63&3836.03&731.30&3365.97&681.39&3846.12&731.89&3506.36\\
49192&953.37&1263.15&998.24&1319.17&920.12&1405.05&1000.61&1249.27\\
33590&1099.50&853.73&1088.25&985.95&1079.74&934.59&1090.61&913.865\\
36666&1286.68&2082.84&1305.37&1831.14&1256.46&2023.56&1303.31&1841.45\\
2154&1478.68&1431.84&1491.54&1456.05&1478.78&1337.21&1487.33&1487.28\\
33908&1770.03&2676.22&1791.54&2716.21&1765.29&2720.40&1787.33&2708.57\\
35666&2144.78&1425.01&2091.54&1695.95&2103.87&1528.35&2087.33&1663.64\\
4971&2378.40&1782.36&2381.95&1707.86&2349.44&1734.95&2369.79&1699.11\\
Total&$\backslash$&15351.22&$\backslash$&15078.30&$\backslash$&15530.26&$\backslash$&15069.54\\
\hline
\hline
\end{tabular}
}
\end{table}
\begin{figure}[hbt!]
\centering
\includegraphics[width=0.5\textwidth, height = 2.3in]{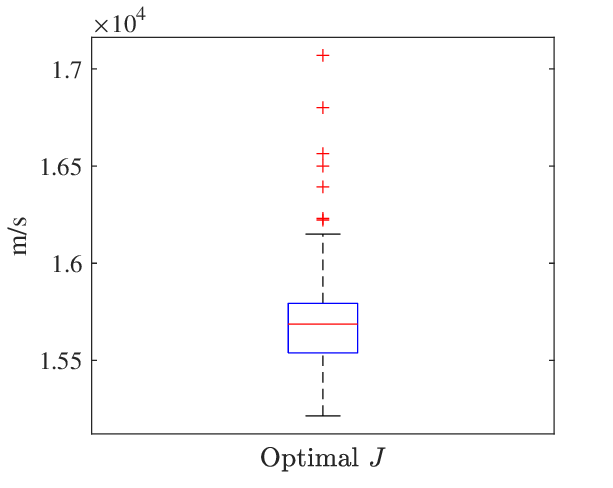}
\caption{Box plot of results obtained by different initial values.}
\label{fig:fig6}
\end{figure}

\begin{figure}[hbt!]
\centering
\includegraphics[width=0.5\textwidth, height = 2.5in]{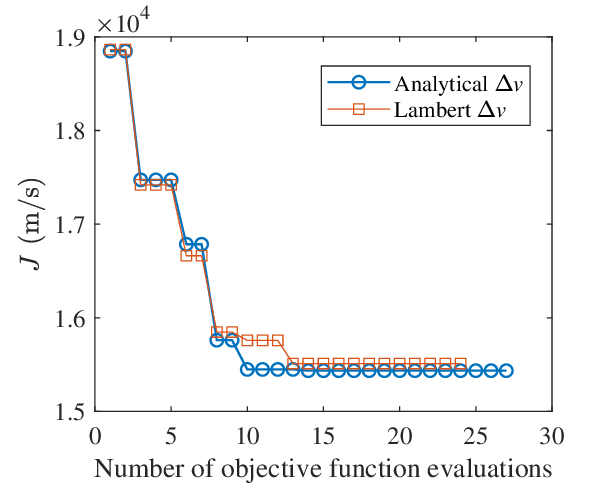}
\caption{Convergence of $J_r$ via the nonlinear programming algorithms.}
\label{fig:fig7}
\end{figure}

\begin{figure}[hbt!]
\centering
\includegraphics[width=0.5\textwidth, height = 2.4in]{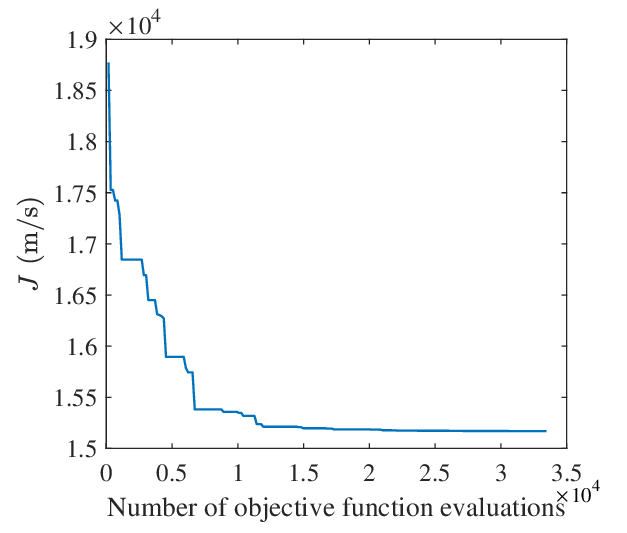}
\caption{Convergence of $J_r$ via DE.}
\label{fig:fig8}
\end{figure}

\begin{figure}[hbt!]
\centering
\includegraphics[width=0.5\textwidth, height = 2.5in]{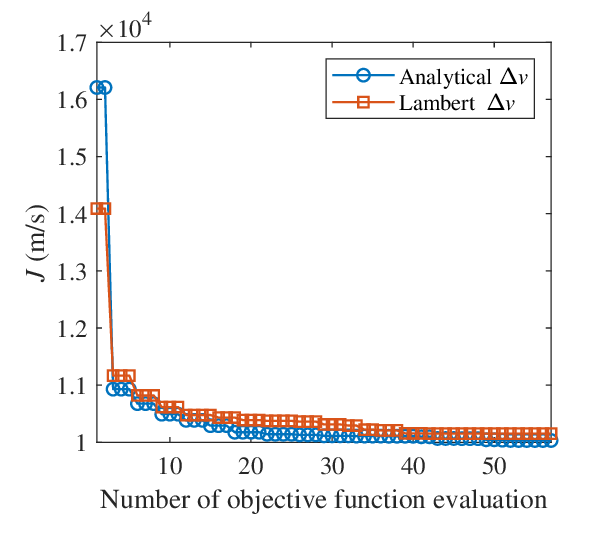}
\caption{Convergence of $J_f$ via the nonlinear programming algorithms.}
\label{fig:fig9}
\end{figure}

\begin{table}[hbt!]
\caption{\label{tab:table5} Optimal solutions of multi-target flyby mission}
\centering
\resizebox{\linewidth}{!}{
\begin{tabular}{lccccccccc}
\hline
\hline
\multirow{2}{*}{Asteroid ID} &\multicolumn{2}{c}{Optimal solution by SQP}	&\multicolumn{2}{c}{Optimal solution by DE}&\multicolumn{2}{c}{Optimal solution by SQP (Lambert)}	&\multicolumn{2}{c}{Optimal solution by DE (Lambert)}\\
\cline{2-9}
&Arrive time (d) &$\Delta v$ (m/s) &Arrive time (d) &$\Delta v$ (m/s) &Arrive time (d) &$\Delta v$ (m/s) &Arrive time (d) &$\Delta v$ (m/s)\\\hline
12095&546&$\backslash$&546&$\backslash$&546&$\backslash$&546&$\backslash$\\
3506&735.35&2328.26&735.62&2328.48&734.63&2260.5&734.81&2260.81\\
49192&1004.52&1348.08&1004.78&1346.73&1004.72&1580.55&1004.94&1579.39\\
33590&1084.52&833.30&1084.78&835.10&1084.72&648.8&1084.94&650.55\\
36666&1309.78&1230.22&1310.17&1228.53&1304.74&1271.49&1305&1269.98\\
2154&1494.81&635.43&1495.12&636.02&1493.61&629.49&1493.92&629.76\\
33908&1780.31&2655.94&1780.93&2656.45&1780.74&2656.4&1781.23&2656.91\\
35666&2110.92&23.47&2112.36&22.52&2084.68&76.7294&2085.49&76.49\\
4971&2367.72&976.59&2350.54&977.27&2334.57&1027.33&2335.05&1027.34\\
Total&$\backslash$&10031.30&$\backslash$&10031.15&$\backslash$&10151.29&$\backslash$&10151.25\\
\hline
\hline
\end{tabular}
}
\end{table}
Next, the multi-target flyby problem with the same targets is also tested. The optimal $J_f$ is 10,031.30 m/s, less than the multi-target rendezvous problem. Table 5 shows the results using different $\Delta v$ evaluation methods and optimization algorithms, which are also similar. Fig. 9 indicates that the convergence requires less than 50 objective function evaluations. The calculation time needed for the proposed method is less because the calculations of analytical $\Delta v$ and its gradients are much more efficient than that of the Lambert's solution. Therefore, the method proposed in this Note is more practical for use as an inner function of published global optimization algorithms \cite{izzo2015, huang2022, zhang2023} to quickly evaluate the mission cost of different sequences with different targets.
\section{Conclusion}
The proposed nonlinear programming model for multi-target rendezvous and flyby missions applies an analytical $\Delta v$ approximation model based on the linear relative motion equations to derive the gradients of constaints and objective functions. Simulation of different asteroid transfers in the main belt indicates that compared with the Lambert's solution, the calculation of the proposed $\Delta v$ approximation is reduced to less than 17\%, and the mean relative error is within 5\%. The analytical gradient-based nonlinear programming algorithm for multi-target sequences can obtain similar results with less time consumption compared with existing evolutionary algorithms and nonlinear programming models. The proposed method can be applied to the preliminary analysis of multi-target missions with pre-determined order and help existing global optimization algorithms improve the efficiency of objective function evaluation.  

\section*{Funding Sources}
The work was supported by the National Natural Science Foundation of China (No. 12202504, 12125207, and 12222213).

\bibliography{main}

\end{document}